\documentclass[aps,prl,reprint]{revtex4-1}  
\usepackage{graphicx}  
\usepackage{hyperref}

\usepackage{todonotes}

\usepackage{floatrow}
\usepackage[caption=false]{subfig}
\newcommand{\aver}[1]{\ensuremath{\left\langle#1\right\rangle}}

\newcommand{\beq}{\begin{equation}}
\newcommand{\eeq}{\end{equation}}
\newcommand{\bea}{\begin{eqnarray}}
\newcommand{\eea}{\end{eqnarray}}
\newcommand{\p}{\partial}
\newcommand{\mb}{\mathbf}

\begin{document}

\title{Fractional Transport in Strongly Turbulent Plasmas}

\author{Heinz Isliker, Loukas Vlahos}
\affiliation{Department of Physics,
 Aristotle University of Thessaloniki\\
GR-52124 Thessaloniki, Greece
   }


\author{Dana Constantinescu}
\affiliation{Department of Applied Mathematics, University of Craiova, Romania}


\date{\today}
\begin{abstract}
We analyze statistically the energization of particles in a large scale environment of strong turbulence
that is fragmented into a large number of distributed 
current filaments.
The turbulent environment is generated through strongly perturbed, 3D,  resistive MHD simulations, 
and it emerges naturally from the nonlinear evolution,
without a specific reconnection geometry being set up.
Based on test-particle simulations,
we estimate the transport coefficients in energy space for use in the classical Fokker-Planck (FP) equation, and we show that the latter fails to reproduce the simulation results. The reason is that transport in energy space is highly anomalous (strange), the particles perform Levy flights, and the energy distributions show extended power-law tails.
Newly then, we motivate the use and derive the specific form of a 
fractional transport equation (FTE), we determine its parameters and the order of the 
fractional derivatives from the simulation data, and we show that the 
FTE is able to reproduce the high energy part of the simulation data very 
well. The procedure for determining the FTE parameters also makes clear that it is the analysis of the simulation data that allows to make the decision whether a classical FP or a FTE is appropriate.

\end{abstract}

\maketitle



\paragraph{Introduction.} 


Particle transport in weakly turbulent environments ($\delta B/B<<1$) has been discussed
extensively for several decades with the use of the Fokker-Planck (FP) equation, mostly
in combination with the quasi-linear (QL) approximation \citep{Kulsrud71,Achterberg81,Petrosian12} but also with other
physical motivations for the form of the transport coefficients \citep[e.g.][]{Morales74}. 

Recent research on the development of \textit{strong magnetic turbulence} 
($\delta B/B \approx 1$) has shown the importance of two scenarios:
Initially extended current filaments (CF) or current sheets (CS) or multiple interacting CF or CS 
develop on fast time scales into 
a strongly turbulent 
environment that is fragmented into a collection of small scale 
current structures
\citep{Matthaeus86,Drake06,Onofri06,Cargill12}. 
On the other hand, reconnection at existing CS is reinforced and new CF or CS are formed by
Alfv\'en waves 
that propagate along complex magnetic topologies 
\citep[see][]{Biskamp89,Lazarian99,Dmitruk04,Arzner04,Hoshino12,Gordovskyy14}.  
In this context, two fundamental questions remain open: 
 (1) Is the FP equation still
	valid in strongly turbulent environments ? (2) How to model transport when the FP
	approach is not valid anymore ?

In this article, 
we consider 
a large scale environment of strong turbulence
that is fragmented into a large number of distributed current filaments, 
and we analyze statistically the energization of particles in this environment, focusing on the case of 
acceleration by the electric field and on the high energy part (tail) of the energy distribution. 

\paragraph{The MHD turbulent environment.}
We consider a strongly turbulent environment as it naturally results from the nonlinear evolution of the MHD equations, in a similar approach as \cite{Dmitruk04}. Thus, we do not set up a specific geometry of a reconnection environment or prescribe a collection of waves as turbulence model, but allow the MHD equations themselves to build 
naturally correlated field structures (which are turbulent, not random) and
coherent regions of intense current densities (current filaments or CS).

The 3D, resistive, compressible and normalized MHD equations used here are 
\beq
\p_t \rho = -\nabla \cdot \mathbf{p}
\eeq
\beq
\p_t \mathbf{p} = 
- \mathbf{\nabla}  \cdot
\left( \mathbf{p} \mathbf{u} - \mathbf{B} \mathbf{B}\right)
-\nabla P - \nabla B^2/2 
\eeq
\beq
\p_t \mathbf{B} = 
-  \nabla \times \mathbf{E}
\eeq
\beq
\p_t (S\rho) = -\mathbf{\nabla} \cdot \left[S\rho \mathbf{u}\right]
\eeq
with $\rho$ the density, $\mathbf{p}$ the momentum density,
$\mathbf{u} = \mathbf{p}/\rho$,
$P$ the thermal pressure,
$\mathbf{B}$ the magnetic field,
$\mathbf{E}   = -  \mathbf{u}\times \mathbf{B} + \eta \mathbf{J}$
the electric field,  
$\mathbf{J} =  \mathbf{\nabla}\times\mathbf{B}$
the current density, $\eta$ the resistivity,
$S=P/\rho^\Gamma$ the entropy,
and $\Gamma=5/3$ the adiabatic index. 

The MHD equations are solved numerically in Cartesian coordinates with the pseudo-spectral method \cite{Boyd2001}, combined with the strong-stability-preserving Runge Kutta scheme of \cite{Gottlieb98}, and by applying periodic boundary conditions to a grid of size $128\times 128\times 128$. As initial conditions we use a superposition of Alfv\'en waves, with a Kolmogorov type spectrum in Fourier space, together with
a constant background magnetic field $B_0$ in the 
$z$-direction. The mean value of the initial magnetic perturbation is $<b> = 0.6 B_0$, its standard deviation is $0.3 B_0$, and the maximum equals $2B_0$, so that we indeed consider strong turbulence. The initial velocity field is 0, and the initial pressure and energy are constant.

For the MHD turbulent environment to build, we let the MHD equations evolve 
until the largest velocity component starts to exceed twice the Alvf\`en speed.
The magnetic Reynolds number at final time is $<|\mathbf{u}|>l/\eta = 3.5\times 10^3$, with $l\approx 0.01$ a typical eddy size,
and the ratio of the energy carried by the magnetic perturbation to the kinetic energy is $(0.5 <b^2>) /(0.5<\rho \mathbf{u}^2>) = 1.4$, which is a second indication that we consider strong turbulence.

The test-particle are tracked in a fixed snapshot of the MHD evolution, 
and we evolve the particles for short times, so we do not probe the scattering of particles off waves, but the interaction with electric fields.
Also, we take into account anomalous resistivity effects by 
increasing the resistivity to $\eta_{an} = 1000 \eta$ locally when
the current density $J=|\mb{J}|$ exceeds a threshold $J_{cr}$.  
The threshold is determined from the frequency distribution of the current density, which exhibits an exponential tail, and the threshold is chosen 
as the value above which the tail is formed ($J_{cr}=60$).
Physical units are  introduced by using the parameters $L=10^5\,$m for the box-size, $v_A=2\times10^6\,$m/s
for the Alfv\'en speed, and $B_0=0.01\,$T for the background magnetic field.
We apply a cubic interpolation of the fields at the grid-points to the actual particle positions.

\paragraph{Test-particle simulations.}
The relativistic guiding center equations (without collisions) are used
for the evolution of the position $\mb{r}$ and the parallel component $u_{||}$ of the relativistic 4-velocity of the particles  \cite{Tao07}, 
\beq
\frac{d\mb{r}}{dt}= \frac{1}{B_{||}^*} 
\left[ \frac{u_{||}}{\gamma} \mb{B}^* + \hat{\mb{b}}\times 
\left(\frac{\mu}{q\gamma}\nabla B -\mb{E}^* \right) \right]
\eeq
\beq
\frac{du_{||}}{dt} = - \frac{q}{m_0 B_{||}^*}\mb{B}^* 
\cdot\left(\frac{\mu}{q\gamma} \nabla B -\mb{E}^* \right)
\eeq
where 
$\mb{B}^*=\mb{B} +\frac{m_0}{q} u_{||}\nabla\times\hat{\mb{b}}$, 
$\mb{E}^*=\mb{E} -\frac{m_0}{q} u_{||} \frac{\p\hat{\mb{b}}}{\p t}$, 
$\mu   = \frac{m_0 u_\perp^2}{2 B}  $
is the magnetic moment, 
$\gamma=\sqrt{1+\frac{u^2}{c^2}}$,
$B=|\mb{B}|$, $\hat{\mb{b}}=\mb{B}/B$, 
$u_\perp$ is the perpendicular component of the relativistic 4-velocity,
and $q$, $m_0$ are the particle charge and rest-mass, respectively.


\begin{figure}[h]
	\centering
	\includegraphics[width=0.90\columnwidth]{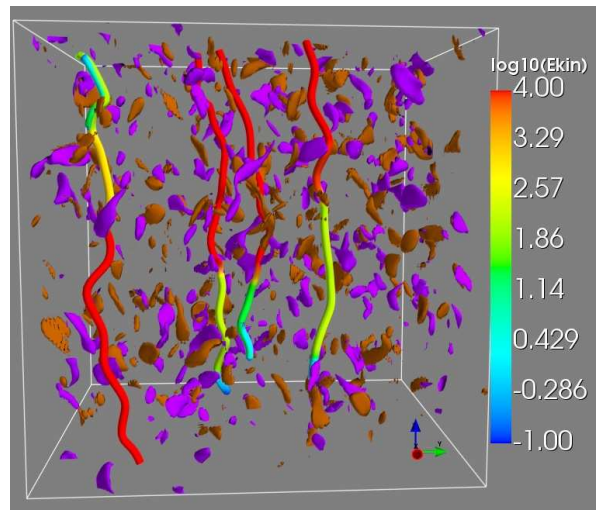}
	\caption{Iso-contours of the supercritical current density component $J_z$ (positive in brown, negative in violet), and a few orbits of energetic particles, colored according to the logarithm of their kinetic energy in keV (see colorbar).}
	\label{fig:snapshot}
\end{figure}

  \begin{figure*}[ht]
	\sidesubfloat[]{\includegraphics[width=0.35\columnwidth]{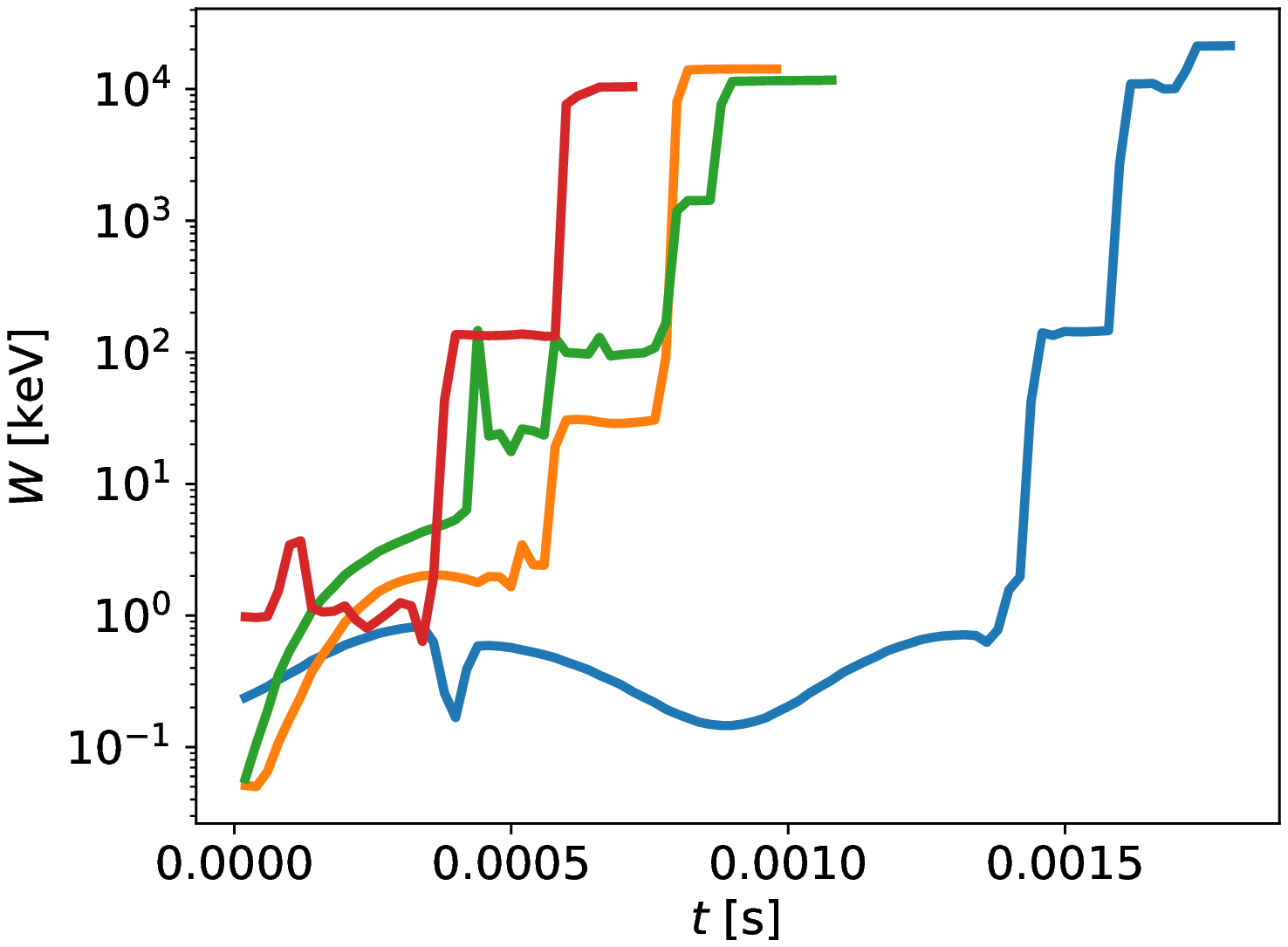}%
		\label{f2:Emean}}\hfill%
	\sidesubfloat[]{\includegraphics[width=0.35\columnwidth]{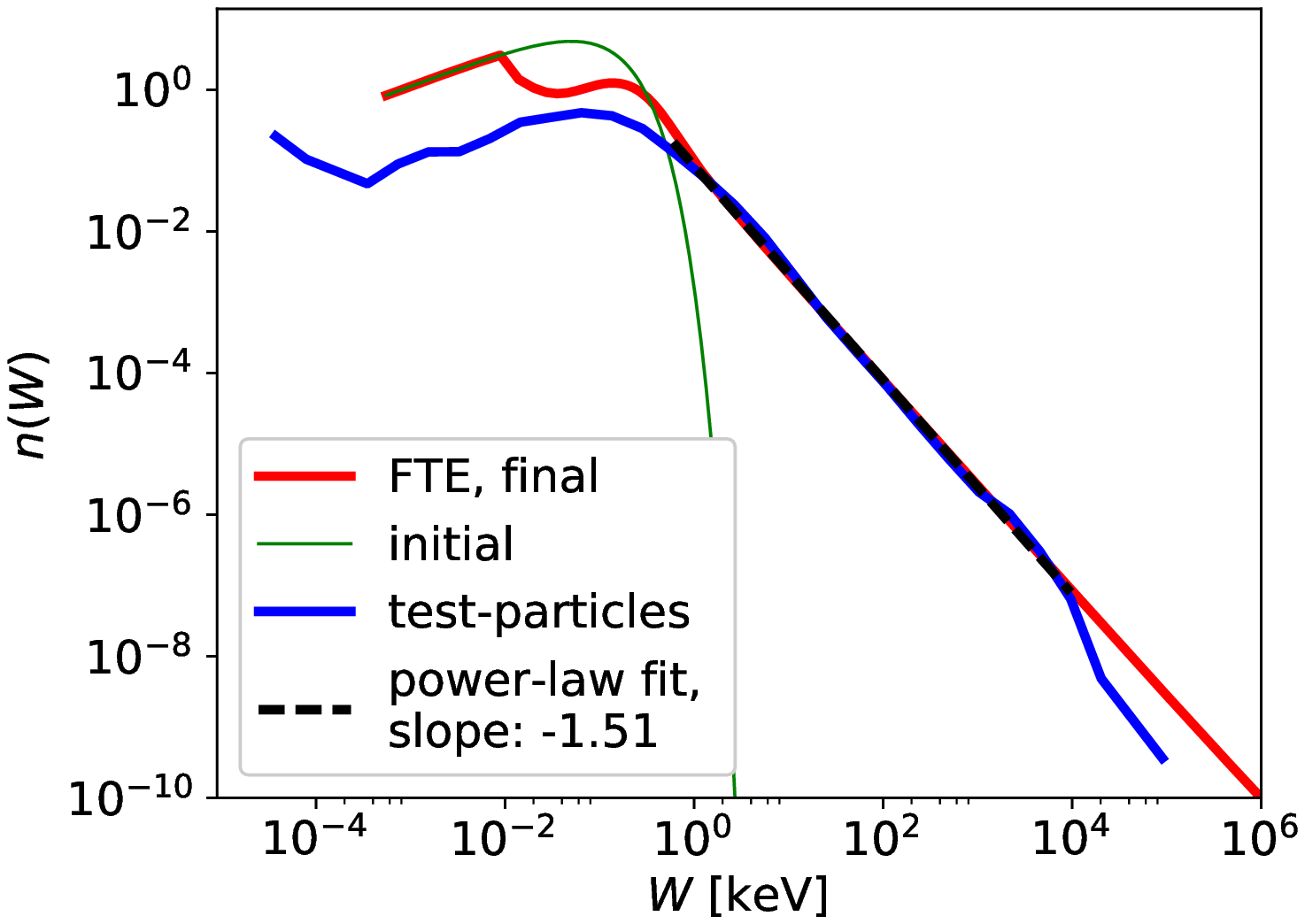}%
		\label{f2:Ekin}}\hfill%
	\sidesubfloat[]{\includegraphics[width=0.35\columnwidth]{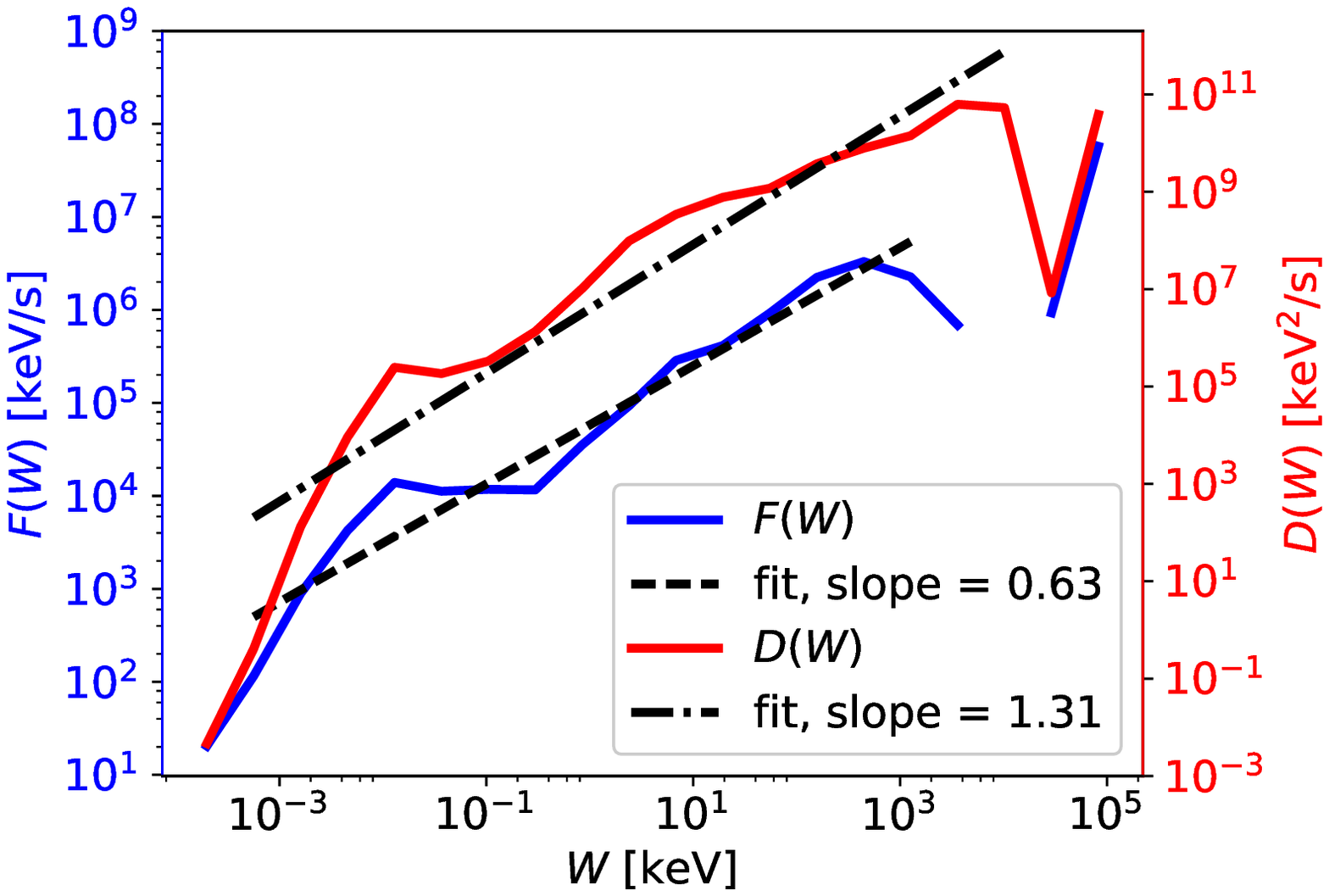}%
		\label{f2:FD}}\hfill%
	\sidesubfloat[]{\includegraphics[width=0.35\columnwidth]{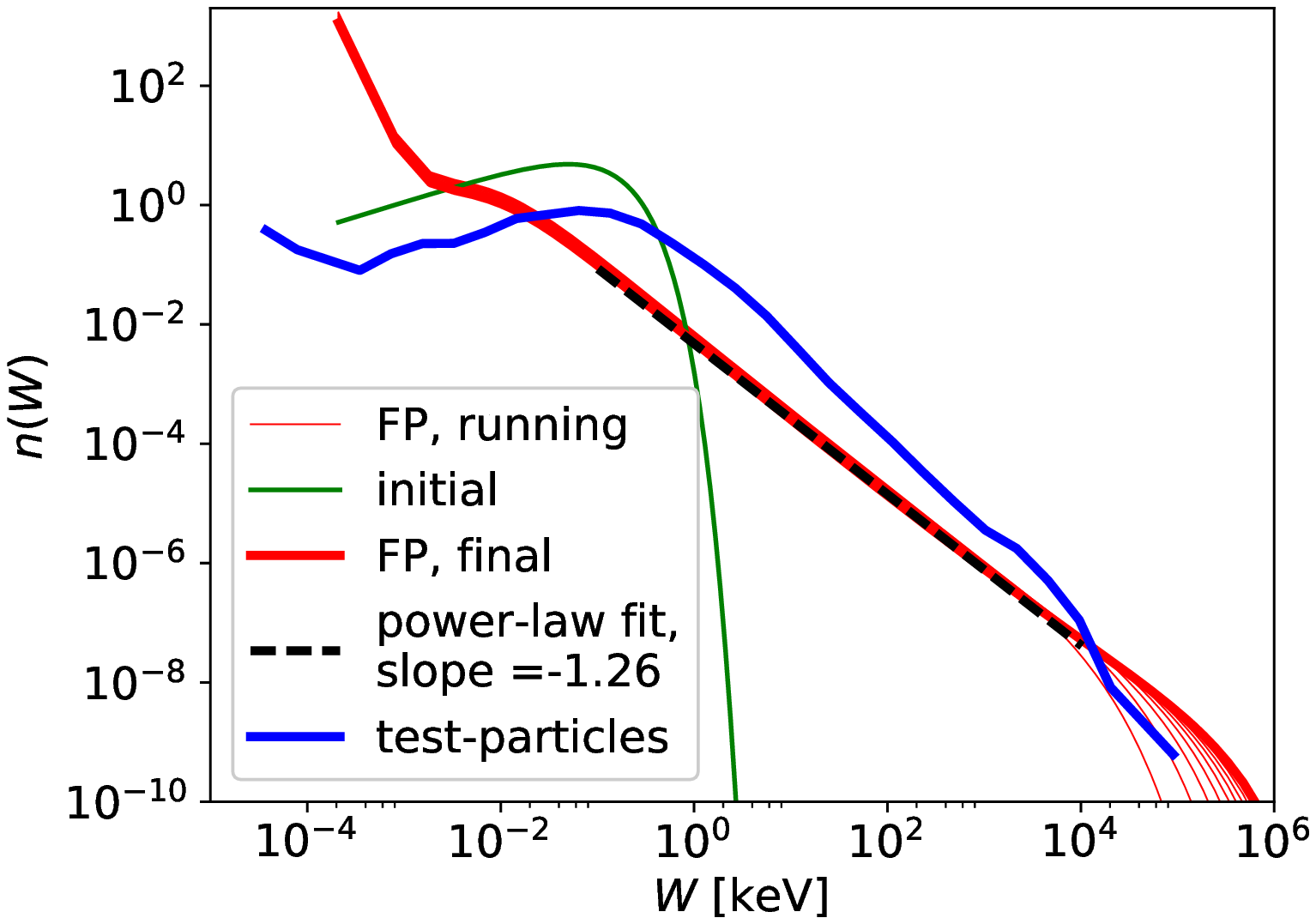}%
		\label{f2:FP}}\hfill%
	\sidesubfloat[]{\includegraphics[width=0.35\columnwidth]{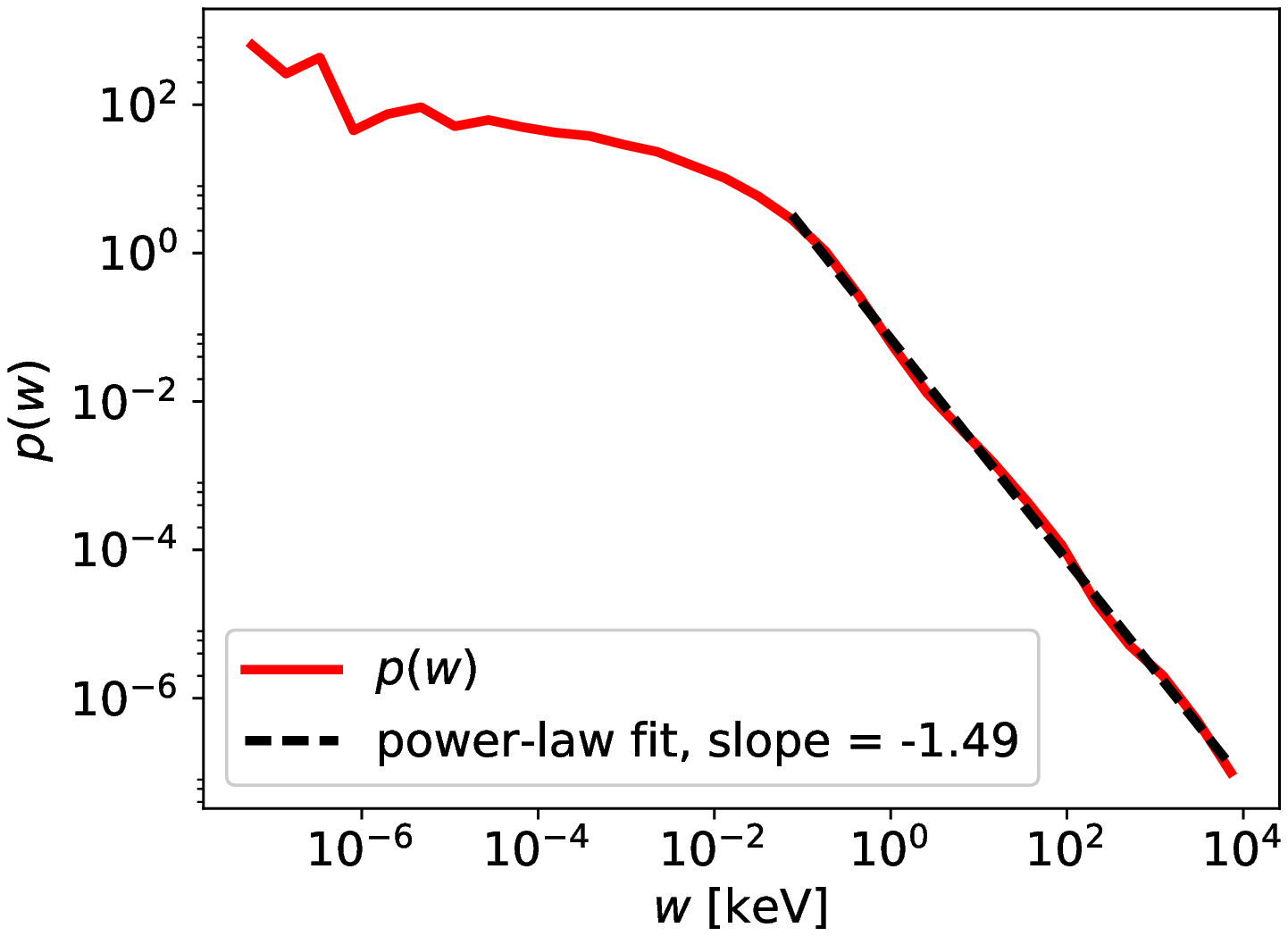}%
		\label{f2:dW}}\hfill%
	\sidesubfloat[]{\includegraphics[width=0.35\columnwidth]{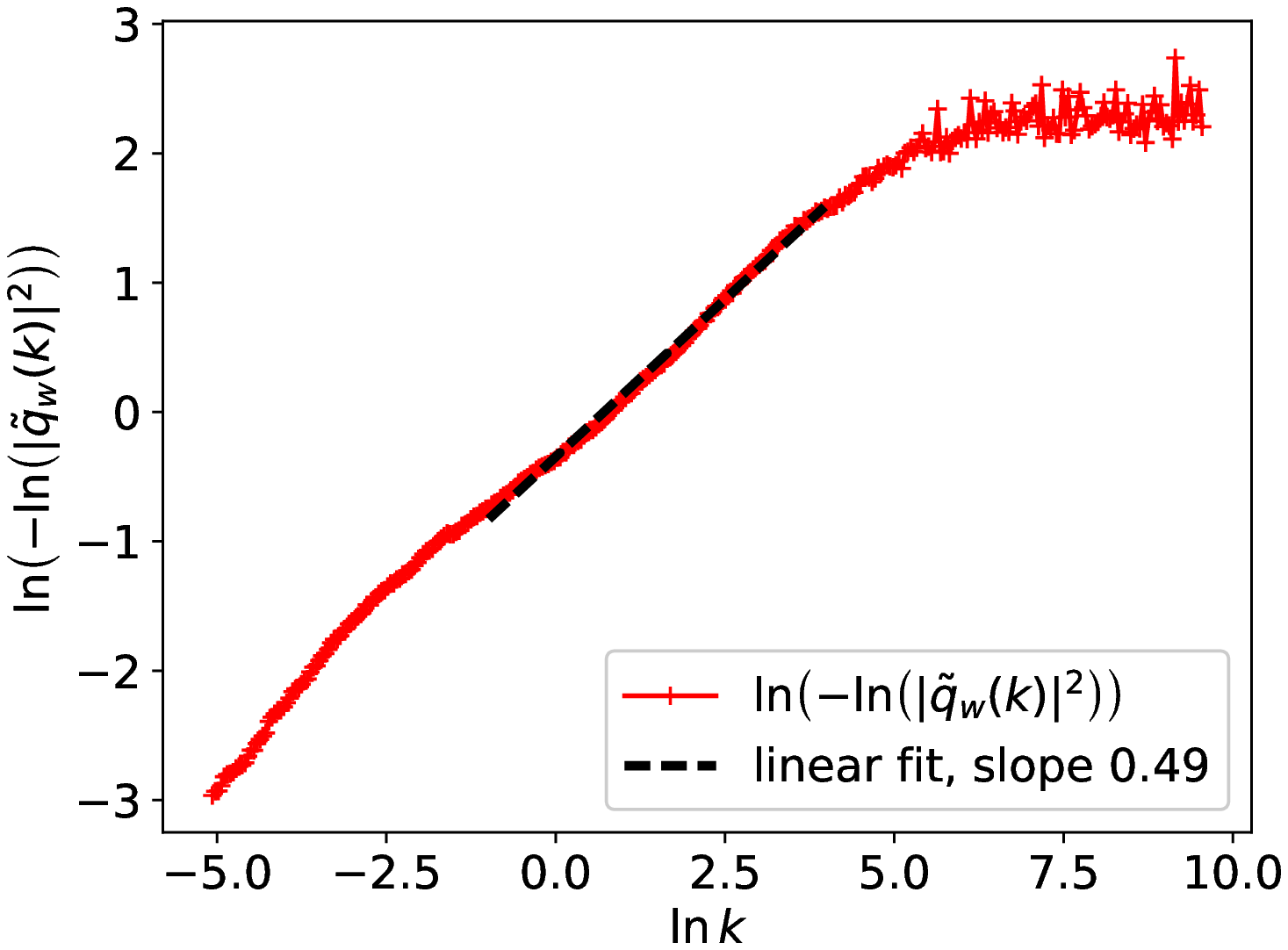}%
		\label{f2:cf}}\hfill%
	\caption{\textit{
			\protect\subref{f2:Emean} The energy evolution of
			4 energetic particles (marked with different colors) is shown.
			\protect\subref{f2:Ekin} Initial and final (at $t
			= 0.002\,$sec) kinetic energy distribution from the test-particle simulations, together with a power-law fit, and the solution of the fractional transport equation at final time.
			\protect\subref{f2:FD} The energy convection and diffusion
			coefficients as a function of the kinetic energy at $t = 0.002\,$sec.
			\protect\subref{f2:FP}
			Solution of the classical FP equation up to a final time of  
			$0.002\,$s, together 
			with the solution at a few intermediate times, and the energy distribution from the test-particle simulations at $t=0.002\,$s.
			\protect\subref{f2:dW}  The distribution $p(w)$ of the energy
			increments $w$ of the particles. 
			\protect\subref{f2:cf} The estimator $\tilde q_w$ of the
			characteristic function $\hat q_w$, based on the sample of 
			energy increments of the test-particle simulations.
	}}\label{f2}
\end{figure*}

The test-particles we consider throughout are electrons.
Initially, all particles are located at random positions, they obey a 
Maxwellian distribution 
with temperature $T=100\,$eV. The simulation box is open, the particles can escape from it when they reach any of its boundaries.

The acceleration process, is very efficient, and we consider a final time
of $0.002\,$s ($7\times 10^5$ gyration periods), at which the asymptotic state has already been reached.
Fig.\ \ref{fig:snapshot} shows the component $J_z$ in the regions of above-critical current density, which clearly are fragmented into a large number of small-scale 
coherent structures within the nonlinear, super-Alfv\'enic MHD environment. The figure also shows the trajectories of 4 particles that reach high energies ($10\,$MeV), 
and Fig.~(\ref{f2:Emean}) shows 
the energy of
the same 4 particles as a function of time.
The particles can lose energy, yet they mostly gain energy in a number of sudden jumps in energy, the energization process thus is localized and there is multiple energization at different current filaments, as is also visible from the color-coded energy in Fig.\ \ref{fig:snapshot}.
Fig.~(\ref{f2:Ekin}) shows the energy distribution at final time, which
exhibits a clear power law part in the intermediate to high
energy range with power-law index $-1.51$, with a slight turnover at the highest energies. 

\paragraph{Transport coefficients and classical FP equation.}
We now turn to the question whether the test-particle results can be reproduced as a solution of the FP equation.
In order to simplify the FP equation (e.g.\ \cite{Gardiner09}), we 
neglect 
the spatial diffusion 
and consider particle diffusion only in energy space, 
including an escape term,
\begin{equation} 
\frac{\partial n}{\partial t}+\frac{\partial }{\partial W} \left [ F n -\frac{\partial [D n]  }{\partial W} \right ]=
-\frac{n}{t_{esc}} ,
\label{diff}
\end{equation}
with $n$ the distribution function, $W$ the kinetic energy, and $t_{esc}$ the escape time, and 
where 
$D$ is the energy diffusion coefficient,
\begin{equation}
D(W,t) =\frac{\aver{\left(W(t+\Delta t)- W(t)\right)^2}_W}{2\Delta t} ,
\label{eq:DWW}
\end{equation}
and $F$ is the energy convection coefficient,
\begin{equation}
F(W,t) =\frac{\aver{W(t+\Delta t)- W(t)}_W}{\Delta t} ,
\label{eq:FW}
\end{equation}
with $\Delta t$ a small time-interval.
With $\aver{...}_W$ we denote the conditional average that $W(t)=W$
(see e.g.\ \cite{Ragwitz2001}).
For the estimate of the coefficients, 
we follow the method described in \cite{Vlahos16},
i.e.\ we keep track of the energy
of the particles at a number of monitoring times separated by
$\Delta t$, and
to account for the conditional averaging, 
we divide the energies
of the particles at time $t$ into a number of logarithmically
equi-spaced bins and perform the requested averages separately
for the particles in each bin.

The estimates of $F(W)$ and $D(W)$ 
at $t
= 0.002\,$s and as function of the energy
are shown in Fig.\ \ref{f2:FD}, and they 
show
a power-law shape, 
with power-law
indexes $a_F = 0.63$ and $a_D = 1.31$.


In \cite{Vlahos16},  we have shown that the described procedure of estimating 
the transport coefficients is consistent
for the case of classical Fermi acceleration, in the sense that the test-particle distribution is reproduced quite well by the FP solution when
the coefficients are inserted into the FP equation and the latter is solved  numerically. We thus
also here insert
$F$ and $D$,
into the FP equation 
and
solve it numerically in the 
energy interval $[0,\infty)$,  
with the method described in \cite{Vlahos16}
(pseudospectral method
based on 
rational Chebyshev polynomials,
combined with the 
backward Euler scheme,
see e.g.\ \cite{Boyd2001}). 
We also include the escape term, for which
the escape time is estimated by assuming that the number of particles that stays inside the simulation box decays exponentially, which yields 
$t_{esc} = 0.004\,$s.

The resulting evolution of the energy distribution up to a final time
of 
$0.002\,$ s, 
is shown in Fig.~(\ref{f2:FP}). 
A clear power-law tail has developed, much flatter though than the one of the test-particle simulations. The discrepancy between the solution of the FP equation and the simulation data actually persists when applying larger integration times.


The failure of the classical FP equation can be explained 
by analyzing  
  the sample of energy increments
  $w_j := W_j(t+\Delta t)- W_j(t)$ 
  (with $j$ the particle index),
  on which the estimates of $F$ and $D$ are
  based (Eqs.~(\ref{eq:FW}) and (\ref{eq:DWW})). 
  These increments follow a power law distribution with index $-1.49$ 
   at intermediate to high energies, 
  as shown in Fig.\ \ref{f2:dW}, and as a
  consequence the particles occasionally perform very large jumps in
  energy space (Levy flights), as illustrated in
  Fig.~\ref{f2:Emean}. 
  The fact that the energy
  increments have a power-law distribution with the specific index
  has several consequences: (1) The estimates of $F$ as a mean  value and $D$ as a variance
  theoretically are infinite, and thus in practice they are very problematic.
  (2) 
  The mean 
  is 
  not representative for a
  scale-free power-law distribution. (3) The prerequisites for deriving a FP equation are not fulfilled, 
  see the comments below.

\paragraph{Fractional transport equation (FTE).}
A general description of transport in energy space is given by
a variant of the Chapman-Kolmogorov equation 
\bea
n(W,t) &=& \int dw\int_{0}^{t} d\tau\,n(W-w,t-\tau) \, q_{w }(w) 
\, q_\tau(\tau) \nonumber \\
&&+ n(W,0)\int_{t}^{\infty}q_\tau(\tau) d\tau
\label{chapkol}
\eea
see e.g.\ \cite{Klafter87,Klages08},
which expresses a conservation law in energy space, and which can be interpreted as describing a Continuous Time Random Walk (CTRW) process. $q_w$ is the probability density for a particle to make a random walk step $w$ in energy,
and $q_\tau$ the probability density for this step to be performed in a time 
interval $\tau$, and for simplicity we have assumed the two probabilities to be independent. When both $q_w$ and $q_\tau$ are practically bounded, 
allowing only small increments, as e.g.\ if they are Gaussians, then the FP equation can be derived from Eq.\ (\ref{chapkol}) through Taylor-expansions
(see e.g.\ \cite{Gardiner09}). For now, we do not make this smallness assumption.

A Fourier Laplace transform ($ W\to k $, $t\to s$), by applying the respective convolution theorems, yields
\beq
\tilde{\hat n}(k,s) = \tilde{\hat n}(k,s) \, \hat q_{w }(k) 
\, \tilde q_\tau(s) + \hat n(k,0) \frac{1 - \tilde q_\tau(s)}{s} 
\label{chapkolFL}
\eeq
which is the Montroll-Weiss equation \cite{Montroll65,Klafter87}, written in non-standard form.

For the distribution of energy increments, expressed in Fourier space (i.e.\ the characteristic function), we 
consider the symmetric stable Levy distributions
$\hat   q_w(k) = \exp(-a|k|^\alpha)$,
with $0<\alpha\leq 2$, which exhibit a power-law tail in energy-space,
$q_w(w)\sim 1/w^{1+\alpha}$ for $\alpha<2$ and $w$ large, and 
for $\alpha=2$ 
they are Gaussian distributions \cite{Hughes95}.
For the waiting time distribution, we assume one sided stable Levy distributions,
expressed in Laplace space,
$\tilde q_\tau = \exp(-bs^\beta)$
with $b>0$ and $0<\beta\leq 1$, which have a power-law tail, 
$q_\tau \sim 1/\tau^{1+\beta}$ for $\beta<1$ and $\tau$ large,
and for $\beta = 1$ they equal $q_\tau(\tau)=\delta(\tau-b)$ \cite{Hughes95}. 

In order to derive a meso-scopic equation, we consider the fluid-limit where  
$w,\,\tau$ are large, and thus $k,\,s$ are small, (e.g.\ \cite{Klages08}, and references therein)
so that the distributions of increments can be approximated as
$\hat   q_w \approx 1-a|k|^\alpha$ and
$\tilde q_\tau \approx 1 - bs^\beta$.
Upon inserting into Eq.\ (\ref{chapkolFL}), we find
\beq
bs^\beta \tilde{\hat n}(k,s) - bs^{\beta-1} \hat n(k,0)  = - a|k|^\alpha \tilde{\hat n}(k,s)   
\label{chapkolFL_fl}
\eeq
which can be written as a fractional transport 
equation
\beq
b D_t^\beta n = a D_{|W|}^\alpha n
\label{fract}
\eeq
with $D_t^\beta$ the Caputo fractional derivative of order $\beta$,
defined in Laplace space as
\beq
{\cal L}\left(D_t^\beta n\right) = s^\beta \tilde n(W,s) -s^{\beta -1} n(W,0)
\eeq
and $D_{|W|}^\alpha$ the symmetric Riesz fractional derivative of order $\alpha$,
defined in Fourier space as
\beq
{\cal F}\left(D_{|W|}^\alpha n\right) = -|k|^\alpha \hat n(k,t) 
\eeq
see e.g.\ \cite{Klages08}.
Note that Eq.\ (\ref{fract}) includes the cases of a pure diffusion or convection equation ($\beta=1$ and $\alpha=2$ or 1, respectively). 
From the derivation of the FTE it is clear that
the order of the fractional derivative is given by the 
index of the power-law tail of the distribution of increments, if any,
otherwise, if the mean and variance of the increments are finite, then 
the classical FP equation is appropriate.

 We need to estimate two parameter sets, $\alpha$, $a$ and  $\beta$, $b$. $\alpha$ can be inferred from the index $z$ of the power-law tail of $p_w(w)$
 in Fig.\ \ref{f2:dW} as $\alpha = -z-1 = 0.49$. As second method to determine $\alpha$ and also $a$, we use the characteristic function approach
 \cite{Borak05,Koutrouvelis80}, with the estimator $\tilde{q}_w$ of the characteristic function $\hat{q}_w$ that is based on the sample of increments $\{w_j\}$ from the test-particle simulations,
 \beq
 \tilde{q}_w(k) = \langle e^{ik w_j}\rangle_j
 \eeq
 for a suitable set of $k$-values. If the $w_j$ obey a stable Levy distribution,
 then $\hat   q_w(k) = \exp(-a|k|^\alpha)$ should hold, and a linear fit to 
 $\ln (-\ln |\tilde{q}_w|^2)$ as a function of $\ln k$ will yield $\alpha$ as the slope
 and $\ln(2 a)$ as the intercept with the $y$-axis. 
 Fig.\ \ref{f2:cf} shows $\hat   q_w(k)$, there indeed is a linear range, 
 and we find
 $\alpha = 0.49$ and $a= 0.36$, with the value of $\alpha$ being 
 equal to
 the one inferred from the power-law tail of the increments. 

We have probed the energy increments over a fixed time interval 
$\Delta t$, and thus we considered the waiting time distribution to be $p_\tau(\tau)=\delta(t-\Delta t)$, from which it follows that $\beta=1$ and 
$b=\Delta t$. This approach seems unavoidable if the test-particle data are given in the form of 
time-series, where there is no direct information on the waiting times between scattering events. 
Thus, in the following we consider the transport equation to have an ordinary, first order derivative in time-direction and a fractional derivative in energy direction,
\beq
\p_t n = (a/b) D_{|W|}^\alpha n - n/t_{esc} ,
\label{fracth}
\eeq
where we also have added an escape term.

For the numerical solution of the fractional transport equation, we use the Gr\"unwald-Letnikov definition
of fractional derivatives (e.g.\ \cite{Kilbas06}) in
the matrix formulation 
of \cite{Podlubny09}, and in order to use the same grid-points in $[0,\infty)$ as above for the solution of the classical FP equation, we use 
the derivative scheme given in \cite{Podlubny13} for non equi-distant grid-points. Also, we apply the fractional derivative only above energies of $10\,$eV, being interested here in the evolution of the high energy part, and considering that the FTE in its current form is not an appropriate tool to model low energy phenomena and heating, being motivated and derived here for modeling long tails at the high energy side of the energy distribution. 

Fig.\ \ref{f2:Ekin} shows the solution of the FTE at 
$t=0.002\,$s, and obviously the distribution from the test-particle simulations is 
very well reproduced in what the power-law tail in its entire extent is concerned. 
Varying the anomalous resistivity $\eta_{an}$ from $10\eta$  to $10^4\eta$, we find that
the FTE we have introduced is always appropriate
and successful in reproducing the simulation data (for $\eta_{an} = \eta$ no power-law tail is being formed).

\paragraph{Summary and Discussion.}
  
  It is to note that the statistical analysis 
  of the simulation data, and in particular the analysis of the distribution of increments, 
  plays a crucial role. First of all, this analysis allows to decide whether a classical FP equation or a FTE is appropriate.

  In the case of anomalous transport, 
  the data 
  have to be analyzed statistically more in depth. 
  The order of the fractional derivative is directly related to the index of the power-law tail of the increments and thus it is easy to estimate, and for an estimate of the 
  scale parameter $a$ we have used the characteristic function method.
  After all, the form of the FTE and its parameters, most prominently the order of the fractional derivative, are 
  directly inferred from the simulation data (and thus they are not universal or unique).
  
  Also, we made no effort to model the low energy part of the distribution, which could possibly be achieved by combining the fractional term in the FTE with classical diffusive and convective terms.

\begin{acknowledgments}
We would like to thank the anonymous referees for their constructive criticism.
This work was supported by (a) the national Programme for the Controlled Thermonuclear Fusion, Hellenic Republic, (b) the EJP Cofund Action SEP-210130335 EUROfusion, and (c) the EUROfusion Consortium under grant agreement No 633053.
The sponsors do not bear any responsibility for the content of this work.
\end{acknowledgments}

\bibliographystyle{pasrev4-1}

\begin{thebibliography}{}
\expandafter\ifx\csname natexlab\endcsname\relax\def\natexlab#1{#1}\fi

\bibitem[{{Achterberg}(1981)}]{Achterberg81}
A. Achterberg, Astronomy \& Astrophysics \textbf{97}, 259
(1981).

\bibitem[{Arzner \& Vlahos(2004)}]{Arzner04}
K. Arzner \& L. Vlahos, The Astrophysical Journal Letters \textbf{605}, L69 (2004).


\bibitem[{{Biskamp} \& {Welter}(1989)}]{Biskamp89}
D. Biskamp, \& H. Welter, Physics of Fluids B \textbf{1}, 1964 (1989).

\bibitem{Borak05}
S. Borak, W. H\"ardle, and R. Weron, in Statistical Tools for Finance and Insurance (Springer, 2005), pp. 21–44.

\bibitem[{Boyd(2001)}]{Boyd2001}
J.~P. Boyd, Chebyshev and Fourier spectral methods (Courier
  Corporation, 2001)

\bibitem[{Cargill {et~al.}(2012)Cargill, Vlahos, Baumann, Drake, \&
  Nordlund}]{Cargill12}
P. Cargill, L. Vlahos, G. Baumann, J. Drake \& \AA. Nordlund, 
  Space science reviews \textbf{173}, 223 (2012).

\bibitem[{{Dahlin} {et~al.}(2015){Dahlin}, {Drake}, \& {Swisdak}}]{Dahlin15}
J.~T. Dahlin, J.~F. Drake, \& M. Swisdak,   Physics of Plasmas \textbf{22},
  100704 (2015).

\bibitem[{{Dmitruk} {et~al.}(2004){Dmitruk}, {Matthaeus}, \&
  {Seenu}}]{Dmitruk04}
P. {Dmitruk},  W.H. {Matthaeus},  \& N. {Seenu},   \apj\ \textbf{617}, 667 (2004).

\bibitem[{{Drake} {et~al.}(2006){Drake}, {Swisdak}, {Che}, \& {Shay}}]{Drake06}
J.F. {Drake},  M. {Swisdak}, H. {Che},  \& M.A. {Shay},  \nat\  \textbf{443}, 553 (2006).





\bibitem[Gardiner(2004)]{Gardiner09}
C.W. Gardiner, Handbook of Stochastic Methods, 4th ed.
(Springer; Berlin \& Heidelberg; 2009)

\bibitem{Gordovskyy14}
M. Gordovskyy, P.K. Browning, E.P. Kontar, and N.H. Bian,
A\&A \textbf{561}, A72 (2014)

\bibitem{Gottlieb98}
S.\ Gottlieb, C.-W. Shu, Mathematics of Computation \textbf{67}, 73 (1998)


\bibitem[{{Guo} {et~al.}(2015){Guo}, {Liu}, {Daughton}, \& {Li}}]{Guo15}
F. {Guo}, Y.-H. {Liu}, W. {Daughton},  \& H. {Li},   \apj\ \textbf{806}, 167
(2015).

\bibitem{Hoshino12}
M. Hoshino, Phys. Rev. Lett. \textbf{108}, 135003  (2012).

\bibitem[Hughes(1995)]{Hughes95}
B.D. Hughes,  Random Walks and Random Environments, Volume 1,
Random Walks, Oxford (Clarendon Press, 1995).

\bibitem[Kilbas et al.(2006)]{Kilbas06} 
A.A. Kilbas, H.M. Srivastava, J.J. Trujillo, Theory and Applications of Fractional Differential Equations (Elsevier, Amsterdam, 2006).

\bibitem[Klafter, Blumen \& Shlesinger(1987)]{Klafter87}
J. Klafter, A. Blumen, M.F. Shlesinger,  Phys.\ Rev.\ A {\bf 35}, 3081
(1987).

\bibitem[Klages, Radons, Sokolov(2008)]{Klages08}
R. Klages, G. Radons, I.M. Sokolov, Anomalous Transport: Foundations and Applications (John Wiley \& Sons, 2008). 

\bibitem{Koutrouvelis80}
I.A. Koutrouvelis,   
Journal of the American Statistical Association {\bf 75}, 918 (1980).


\bibitem[{Kulsrud \& Ferrari(1971)}]{Kulsrud71}
R.~M. Kulsrud, A. Ferrari, Astrophysics and Space Science \textbf{12}, 302 (1971).

\bibitem[{{Lazarian} \& {Vishniac}(1999)}]{Lazarian99}
A. {Lazarian},  \& E.T. {Vishniac},   \apj\ \textbf{517}, 700 (1999)



\bibitem[{{Matthaeus} \& {Lamkin}(1986)}]{Matthaeus86}
W.H. {Matthaeus},  \& S.L. {Lamkin},  Physics of Fluids \textbf{29}, 2513 (1986).

\bibitem[Montroll \& Weiss(1965)]{Montroll65}
E.W.\ Montroll, G.H.\ Weiss, J.\ Math.\ Phys.\ {\bf 6}, 167 (1965).

\bibitem[{{Onofri} {et~al.}(2006){Onofri}, {Isliker}, \& {Vlahos}}]{Onofri06}
M. {Onofri}, H. {Isliker}, \& L. {Vlahos},  Physical Review Letters \textbf{96},
  151102 (2006).

\bibitem[{Petrosian(2012)}]{Petrosian12}
V. Petrosian, Space science reviews \textbf{173}, 535 (2012).

\bibitem[Podlubny et al.(2009)]{Podlubny09}
I. Podlubny, A. Chechkin, T. Skovranek, Y.Q. Chen, B.M. Vinagre Jara,
Journal of Computational Physics \textbf{228},  3137  (2009).

\bibitem[Podlubny et al.(2013)]{Podlubny13}
I. Podlubny, T. Skovranek,
B.M. Vinagre Jara, I. Petras, V. Verbitsky, Y.Q. Chen YQ., 
Phil Trans R Soc A \textbf{371}, 20120153 (2013).

\bibitem[{Ragwitz \& Kantz(2001)}]{Ragwitz2001}
M. Ragwitz,  \& H. Kantz,  Physical Review Letters \textbf{87}, 254501 (2001).

\bibitem{Tao07}
X.\ Tao, A.\ A.\ Chan, and Alain J.\ Brizard,
Phys.\ Plasmas {\bf 14}, 092107 (2007).

\bibitem{Vlahos16}
L. Vlahos, Th. Pisokas, H. Isliker, V. Tsiolis, A. Anastasiadis,
The Astrophysical Journal Letters \textbf{827}, L3, (2016).

\bibitem{Morales74}
G. J. Morales and Y. C. Lee, Phys.\ Rev.\ Lett.\ \textbf{33}, 1534 (1974).




\end{thebibliography}

\end{document}